\documentclass[prd,twocolumn,aps,superscriptaddress,eqsecnum,floatfix]{revtex4-1}
\usepackage{empheq}
\usepackage{amsmath}
\usepackage{amssymb}
\usepackage{dsfont}
\usepackage{graphicx}
\usepackage{hyperref}
\usepackage{stackrel}
\usepackage{color}
\usepackage{comment}

\def \ee{\end{equation}}
\def \be{\begin{equation}}
\def \eea{\end{eqnarray}}
\def \bea{\begin{eqnarray}}

\begin{document}

\title{Symmetries of 
relativistic world-lines}

\author{Benjamin Koch}
\email{bkoch@fis.puc.cl}
\affiliation{Pontificia Universidad Cat\'olica de Chile \\ Instituto de F\'isica, Pontificia Universidad Cat\'olica de Chile, \\
Casilla 306, Santiago, Chile}
\author{Enrique Mu\~noz}
\email{munozt@fis.puc.cl}
\affiliation{Pontificia Universidad Cat\'olica de Chile \\ Instituto de F\'isica, Pontificia Universidad Cat\'olica de Chile, \\
Casilla 306, Santiago, Chile}
\author{Ignacio A. Reyes}
\email{ignacio.reyes@physik.uni-wuerzburg.de}
\affiliation{Pontificia Universidad Cat\'olica de Chile \\ Instituto de F\'isica, Pontificia Universidad Cat\'olica de Chile, \\
Casilla 306, Santiago, Chile}
\affiliation{Insitut f{\"u}r Theoretische Physik und Astrophysik, Julius-Maximilians-Universit{\"a}t W{\"u}rzburg, Am Hubland, 97074, Germany}

\begin{abstract}
Symmetries are essential for a consistent formulation of many quantum systems.
In this paper we discuss a previously unnoticed
 symmetry, which is present for any
Lagrangian term that involves $\dot{x}^2$. 
As a basic model that incorporates the fundamental symmetries of
quantum gravity and string theory,
we consider the Lagrangian action of the relativistic point particle. A path integral quantization
for this seemingly simple system
has for long presented notorious problems. 
Here we show that those problems are overcome by taking into account the newly discovered additional symmetry, leading directly to the exact Klein-Gordon propagator. 
\end{abstract}

{\color{blue}
}
\maketitle

\section{Introduction}

Gauge symmetries and the global symmetry of special relativity are the essential ingredients of modern quantum physics,
providing the most fundamental description of nature \cite{Weinberg:1967tq}.
However, up to now, any straightforward attempt to truly unify those concepts has failed.
The most prominent example for this failure is that, a consistent quantum description
of General Relativity, whose gauge symmetry is a local generalization of the global symmetry of special relativity, is still missing.
Similar technical problems arise in the context of String Theory~\cite{Polchinski:book}, mainly due to the square-root kinetic energy
term in the Lagrangian action.
The problem does not seem to be only with General Relativity or String Theory itself, but rather with the unification of 
the fundamental concepts of gauge symmetries and relativity within quantum mechanics. Clearly, a crucial
step towards achieving this unification is the correct analysis of symmetries in the most simple theory with general covariance, the Lagrangian description of relativistic point particle world-lines
whose action is given by
\begin{align}\label{i}
I[x(\lambda)]&=-m\int d\lambda \sqrt{-\dot{x}^\mu \dot{x}^\nu\eta_{\mu\nu}}.
\end{align}
Due to the non-quadratic form of this action,
the path integral formulation of
this apparently basic problem has posed significant difficulties. 
By discovering and applying an additional 
symmetry present in (\ref{i}), we were able to derive an exact explicit expression for the corresponding relativistic propagator, thus filling one of the missing pieces for several of the most pressing problems
in modern theoretical physics.

Even though the Path Integral (PI) \cite{Feynman:1950ir} of \eqref{i} is the natural generalization of its non-relativistic counterpart, it has presented many difficulties. With the advent of Quantum Field Theory (QFT) part of the problems where put aside since one knows what the result should be,
namely the propagator for the free massive scalar field from QFT~\cite{Itzykson:Book}.
We shall work in Minkowski spacetime with signature $\eta=(-,+,\hdots,+)$.
Despite of the simplicity of the action (\ref{i}),
there remain some essential problems with the path integral formulation of this theory. 
As is well known also in statistical mechanics of relativistic particles, the propagator 
 fails to satisfy a naive application of the Chapman-Kolmogorov relation $(x_f|0)=\int d^dx' (x_f|x')(x'|0)$, 
seemingly implying that the theory is not unitary \cite{Kleinert:book}. 
A ``technical'' problem is also relevant: 
A direct evaluation of a path integral with a square root action is 
highly non-trivial \cite{Prugo80,Fuku86}.
Thus, for the case of the relativistic point particle
one is limited to evaluate the path integral of an alternative quadratic action which 
is equivalent to (\ref{i}) at the classical level \cite{Polchinski:book,Kleinert:book,Redmount:1990mi,Fradkin:1991ci},
or to use some other kind of approximation \cite{Padmanabhan:1994}.
This is similar to the situation
in string theory, where the presence of the square root has prevented a direct evaluation of the Nambu-Goto path integral, and the Polyakov action was introduced precisely to surmount these difficulties \cite{Polchinski:book}. 

In summary,
a direct PI calculation of \eqref{i} is still lacking. 
The purpose of this letter is to fill in this gap by a very simple observation: 
``Any Lagrangian term of the form $\dot{x}^2$ has a symmetry that must be accounted for.'' The paper is organized as follows.
In the next section the crucial symmetry aspects of the problem are discussed.
This is followed by an explicit Faddeev-Popov procedure for the PI of \eqref{i},
taking into account those symmetries.
In the final section, we summarize the results and comment
on technical and conceptual implications.

\section{Symmetries of free particles}

 As is well known, the action (\ref{i}) possesses invariance under reparametrizations: if $\lambda\rightarrow \lambda'(\lambda)$ for an arbitrary function $\lambda'(\lambda)$ (we assume monotonic, differentiable, and integrable), the action remains unchanged. For infinitesimal transformations where $\lambda'=\lambda-h(\lambda)$ with $h(\lambda)$ small, 
the induced transformation of the fields are $\delta_r x^\mu=h(\lambda)v^\mu$ where $v^\mu=\dot{x}^\mu=dx^\mu/d\lambda$. 
The subindex ``r'' stands for ``reparametrization''.

From the Hamiltonian point of view, one defines canonical momenta through $p_\mu\equiv \partial L/\partial \dot{x}^\mu=m\dot{x}^\mu/\sqrt{-\dot{x}^2}$, and then reparametrizations are generated by the first class constraint 
\be\label{constrphi}
\phi\equiv p^\mu p_\mu+m^2,
\ee 
via Poisson brackets, 
\bea\label{deltar}
\delta_r x^\mu=\{x^\mu, h \phi\}&=&2hp^\mu,\\ \nonumber
\delta_r p_\nu=\{p_\nu,\phi\}&=&0.
\eea
Let us now examine a different symmetry of this action, which will play a fundamental role in the argument below. Any kinetic term $v^\mu v_\mu$ possesses an additional symmetry: one can locally rotate the velocity to $v'^\mu$ with the constraint that $v'^2=v^2$,  i.e. local $SO(1,d-1)$ rotations of the velocity. This transformation involves $d-1$ arbitrary functions of the parameter, one for each of the angles of the $S^{d-1}$ sphere. We will refer to these as `local velocity rotations'. Infinitesimally, this condition is $v\cdot \delta v=0$, and the most general variation $\delta_o v^\mu$ orthogonal to the velocity is
\begin{align}\label{dov}
\delta_o v^\mu(\lambda)={f}^\mu (\lambda)-(f\cdot v) \frac{v^\mu}{|v|^2}.
\end{align}
The function $f^\mu(\lambda)$ is assumed to 
be well behaved (integrable, differentiable, and monotonic).
Integrating these equations gives the transformation of the fundamental fields $\delta_o x^\mu$.
Thus, the symmetry is local in the velocities, but non-local in the position variables, unlike
usual gauge symmetries. However, the point is that if one factors this symmetry out of the path 
integral, the inconsistency with a naive Chapman-Kolmogorov relation can be solved
and the calculation of the exact propagator results in a straight forward way.

It is instructive to count how many extra degrees of freedom are subtracted from the action,
due to this new symmetry.
The local transformations of $SO(1,d-1)$
contain $d(d-1)/2$ degrees of freedom.
However, those contain the subgroup $SO(1,d-2)$ with
$(d-1)(d-2)/2$ parameters which leave a given velocity vector $v^\mu$ constant, and which are thus acting trivially, so they must not be fixed in the path integral.  
The remaining non-trivial degrees of freedom (which actually change $v^\mu$)
correspond precisely to the $d-1$ transformations that are orthogonal to the velocity.

As it is shown below, by  factorizing this additional symmetry out from the path integral, a standard Faddeev-Popov \cite{Faddeev:1967fc} calculation leads to the correct propagator for the relativistic point particle. \\


\section{Path integral: Faddeev-Popov method}

The object to be computed is
\bea
(x_f|0) = \int_{0}^{x_f}\mathcal{D}x\ e^{iI[x]},
\label{eq_00}
\eea
where we have used the global translation invariance in space-time to set
$x^\mu(\lambda_i)=0$ and $x^\mu(\lambda_f)=x^\mu_f$. This can be rewritten by introducing a Dirac delta identity,
\begin{align}\label{eq_2}
(x_f|0)&=\int_0^{x_f} \mathcal{D}x\ \int_{-\infty}^\infty dS\ e^{iS}\delta\left( S-I[x] \right) \nonumber \\
&=\int_{-\infty}^\infty dS\ e^{iS} \Omega(S)
\end{align}
where the sum over histories is now expressed as an ordinary integral over the values of the action $S$. Here, we have defined the volume or multiplicity 
$\Omega(S)$ of trajectories connecting
the points $0$ and $x_f$, that share the same value of the action $S$ as
\bea
\Omega(S) = \int_{0}^{x_f}\mathcal{D}x(\lambda) \delta\left( S + m\int_{\lambda_i}^{\lambda_{f}}\sqrt{-\dot{x}^{\mu}\dot{x}_{\mu}}d\lambda \right).
\label{eq_3}
\eea
By explicitly computing $\Omega$ we will solve the path-integral defined in (\ref{eq_2}).

We now turn to the Faddeev-Popov procedure for factoring out the redundancy of the PI. In the calculation of $\Omega$, we can start by exploiting reparametrization symmetry by choosing as a convenient parametric scale
the interval (or proper time) of the particle ($c=1$), $d\tau = \sqrt{-dx^{\mu}dx_{\mu}}$. We fix
this choice by inserting the Fadeev-Popov functional identity (see Appendix \ref{AppB})
\begin{eqnarray}
1 &=& \int \mathcal{D}v(\lambda)\delta\left[v(\lambda)^2 + \dot{x}^\mu\dot{x}_\mu \right]\det\left[\frac{\delta \left(v(\lambda)^2 + \dot{x}^\mu\dot{x}_\mu  \right)}{\delta v(\lambda')}  \right]\nonumber\\
&=& \int \mathcal{D}v(\lambda) \delta\left[v(\lambda)^2 + \dot{x}^\mu\dot{x}_\mu \right]\det\left[2\delta(\lambda-\lambda')v(\lambda))\right]
\label{eq_vlambda}
\end{eqnarray}

Notice that, in a given discretization $\lambda \in [\lambda_i,\lambda_f]\rightarrow\left\{\lambda_j \right\}$, the values of the function $v(\lambda)\rightarrow v(\lambda_j) = v_j$, and hence the determinant possesses the simple structure $\det\left[2\delta(\lambda-\lambda')v(\lambda)\right] = \prod_{j}2 v_j$.
Let us define the differential proper time as
\be\label{diffpropt}
d\tau = \sqrt{-dx^{\mu}dx_{\mu}}.
\ee
The invariance of the total interval upon re-parameterization, in differential form
$v(\lambda)d\lambda = d\tau$
implies 
\bea
\int_{\lambda_i}^{\lambda_f} v(\lambda) d\lambda &=& \int_{\tau_i}^{\tau_f} d\tau(\lambda)= \int_{0}^{x_f}\sqrt{-dx^{\mu}dx_\mu}.
\eea
Moreover, the reparametrization introduces, via chain-rule, a multiplicative scale factor
in the velocities:
\begin{eqnarray}
\dot{x}_{\mu}\dot{x}^{\mu} = \left(\frac{d\tau}{d\lambda} \right)^2 \frac{dx_{\mu}}{d\tau}\frac{dx^{\mu}}{d\tau} = v^2(\lambda(\tau))\frac{dx_{\mu}}{d\tau}\frac{dx^{\mu}}{d\tau}
\label{eq_xdot}
\end{eqnarray}
Therefore, upon inserting Eq.(\ref{eq_vlambda}) the volume becomes
\begin{widetext}
\bea
\Omega(S) &=& \int_{0}^{x_f} \mathcal{D}x(\lambda) \int\mathcal{D}v(\lambda) \det\left[2\delta(\lambda-\lambda')v(\lambda)\right] 
\delta\left[ v^2(\lambda) + \dot{x}^2 \right]\delta\left(S + m\int_{\lambda_i}^{\lambda_f}\sqrt{-\dot{x}^2}d\lambda  \right)\nonumber\\ \nonumber
&=& \int \mathcal{D}v(\lambda(\tau))\int_{0}^{x_f}\mathcal{D}x(\tau)
\delta\left[v^2(\lambda(\tau))\left( 1 + \frac{dx^{\mu}}{d\tau}\frac{dx_{\mu}}{d\tau}\right) \right]\det\left[2\delta(\lambda-\lambda')v(\lambda)\right]  \delta\left(S + m\int_{\tau_i}^{\tau_f}d\tau\sqrt{-\frac{dx^{\mu}}{d\tau}\frac{dx_{\mu}}{d\tau}}
 \right)\nonumber\\
 &=& \left[ \int \mathcal{D}v(\lambda(\tau)) \right] \int_{0}^{x_f}\mathcal{D}x(\tau)
\delta\left[ 1 + \frac{dx^{\mu}}{d\tau}\frac{dx_{\mu}}{d\tau} \right] \delta\left(S + m\int_{\tau_i}^{\tau_f}d\tau\sqrt{-\frac{dx^{\mu}}{d\tau}\frac{dx_{\mu}}{d\tau}}
 \right)
 \label{eq_omeg1}
\eea
\end{widetext}
Thus, clearly the (infinite) `volume' $V_r = \int\mathcal{D}v(\tau)$ associated to reparametrization symmetry has factored out. Let us now proceed by renaming the integration variables that define the trajectories in the path-integral, upon defining the ``momenta''
\be
p^{\mu}(\tau) \equiv m \frac{dx^{\mu}}{d\tau}.
\label{eq_pmu}
\ee 
This definition implies the global identity (see Appendix \ref{AppA})
\begin{eqnarray}
\int_{\tau_i}^{\tau_f} p^{\mu}(\tau)d\tau = m\int_{0}^{x_f}dx^{\mu} = m\, x_f^\mu.
\label{eq_pint}
\end{eqnarray}
Notice that, in terms of the momenta, and using the definition of the proper time, 
the action functional acquires the simpler form:
\bea
\frac{S[x]}{m} &=& -\int_{\tau_i}^{\tau_f}  \sqrt{-\frac{dx^{\mu}}{d\tau}\frac{dx_{\mu}}{d\tau}}d\tau 
=- \int_{\tau_i}^{\tau_f}  \sqrt{-\frac{dx^{\mu}}{d\tau}\frac{dx_{\mu}}{d\tau}}^2 d\tau\nonumber\\
&=&  \int_{\tau_i}^{\tau_f} \frac{dx^{\mu}}{d\tau}\frac{dx_{\mu}}{d\tau} d\tau =  \frac{1}{m^2}\int_{\tau_i}^{\tau_f} p^{\mu} (\tau)p_{\mu}(\tau)d\tau,
\label{eq_Sp}
\eea
where we have used the definition of the differential proper time (\ref{diffpropt}).
Thus, changing the integration measure $\mathcal{D}x(\tau) \rightarrow \mathcal{D}p(\tau)$ (see Appendix \ref{AppA}), we have (up to an action-independent Jacobian) the expression
\bea\label{eq28}
\frac{
\Omega(S)}{V_r}&= & \int \mathcal{D}p(\tau)\delta\left[
m^2 + p^2(\tau)
\right] \cdot\\ \nonumber &&
\delta\left(
S - \frac{1}{m}\int_{\tau_i}^{\tau_f}p_{\mu}(\tau)p^{\mu}(\tau)d\tau
\right).
\label{eq_ome}
\eea

Finally, we still need to factor out the local velocity $SO(1,d-2)$ rotations as explained in the introduction. Given an arbitrary constant d-vector $k^{\mu}$ satisfying $k^2 = - m^2$,
there is a unique rotation in  $G = SO(1,d-2)$ that connects it to each d-momenta along the trajectory, i.e. there exists a matrix $\Lambda_{\nu}^{\mu}(\tau)$ such that $p^{\mu}(\tau) = \Lambda_{\nu}^{\mu}(\tau)k^{\nu}$. We can parametrize each rotation, in the vicinity of the identity, by a set
of infinitesimal antisymmetric parameters as $\Lambda_{\nu}^{\mu}(\tau) = \delta_{\mu}^{\nu} + \omega_{\nu}^{\mu}(\tau)$. Therefore, we have the functional identity (see Appendix \ref{AppB})
\begin{eqnarray}
1 &=& \int_{G}\mathcal{D}\Lambda \delta[p^{\mu}(\tau) - 	\Lambda_{\nu}^{\mu}(\tau)k^{\nu}]\\
&&\det\left[\frac{\delta}{\delta\omega_{\nu'}^{\mu'}(\tau')}\left( p^{\mu}(\tau) - 	\Lambda_{\nu}^{\mu}(\tau)k^{\nu}\right) \right]\nonumber\\
&=& \int \mathcal{D}\omega(\tau) \delta[p^{\mu}(\tau) - 	\Lambda_{\nu}^{\mu}(\tau)k^{\nu}]  \Delta[k],
\label{eq_Fad1}
\end{eqnarray}
where we have defined the Fadeev-Popov determinant
\begin{eqnarray}
\Delta[k] = \det\left[\delta(\tau-\tau')\delta^\mu_{\mu'}\delta^\nu_{\nu'}k^{\nu'}\right].
\label{eq_Fad2}
\end{eqnarray}
After Eq.(\ref{eq_Fad1}) and Eq.(\ref{eq_Fad2}), we notice that $\Delta[k]$ is independent of $p^{\mu}(\tau)$. Therefore,
let us now define the constant
\bea
C &=& \int d^{d}k \left(\Delta[k]\right)^{-1}\nonumber\\
&=& \int d^{d}k \int_{G} \mathcal{D}\Lambda \delta[p^{\mu}(\tau) - 	\Lambda_{\nu}^{\mu}(\tau)k^{\nu}].
\label{eq_Const}
\eea
Inserting Eq.(\ref{eq_Const}) into \eqref{eq28}, we obtain
\begin{widetext}
\bea
\Omega(S) &=& \frac{V_r}{C} \int d^{d}k \int_G \mathcal{D}\Lambda \int \mathcal{D}p(\tau) 
\delta\left[p^{\mu}(\tau) - \Lambda_{\nu}^{\mu}(\tau)k^{\nu} \right]\delta\left[m^2 + p^2(\tau) \right]\delta\left( S - \frac{1}{m}\int_{\tau_i}^{\tau_f}p_{\mu}(\tau)p^{\mu}(\tau)d\tau \right)\nonumber\\ 
&=&  \frac{V_r}{C} \int d^{d}k\, \delta\left(m^2 + k^2 \right)\int_G \mathcal{D}\Lambda \int \mathcal{D}p(\tau) 
\delta\left[p^{\mu}(\tau) - \Lambda_{\nu}^{\mu}(\tau)k^{\nu} \right]\delta\left( S - \frac{k_{\mu}}{m}\int_{\tau_i}^{\tau_f}\left[\Lambda^{-1}(\tau)\right]_{\nu}^{\mu}p^{\nu}(\tau)d\tau \right).
\label{eq_omeg3}
\end{eqnarray}
\end{widetext}

Now, let us change the momenta within the path integral by the rotation $p^{\mu}(\tau)\rightarrow \Lambda^{\mu}_{\nu}(\tau)p^{\nu}(\tau) \equiv p_{\Lambda}^{\mu}(\tau)$. 
 Thus, we obtain
\begin{eqnarray}\nonumber
\Omega(S) &=&  \frac{V_r}{C} \int d^{d}k \delta\left(m^2 + k^2\right)\int_G \mathcal{D}\Lambda \int \mathcal{D}p_{\Lambda}(\tau) \ \\
&&\delta\left[p^{\mu}_{\Lambda}(\tau) - \Lambda_{\nu}^{\mu}(\tau)k^{\nu} \right]
\delta\left( S - \frac{k_{\mu}}{m}\int_{\tau_i}^{\tau_f} \left[\Lambda^{-1}\right]_{\nu}^{\mu} p_{\Lambda}^{\nu}d\tau \right)
\nonumber\\ \nonumber
&=&  \frac{V_r}{C} \int d^{d}k \delta\left(m^2 + k^2\right)\int_G \mathcal{D}\Lambda \int \mathcal{D}p(\tau) \ \\ 
&&
\delta\left[\Lambda_{\nu}^{\mu}(\tau)\left(p^{\nu}(\tau) - k^{\nu}\right) \right]
\delta\left( S - \frac{k_{\mu}}{m}\int_{\tau_i}^{\tau_f} p^{\mu}(\tau)d\tau \right)\nonumber\\ \nonumber
&=&  \frac{V_r}{C} \int d^{d}k \delta\left(m^2 + k^2\right)\int_G \mathcal{D}\Lambda \int \mathcal{D}p(\tau) \ \\
&&\delta\left[p^{\mu}(\tau) - k^{\mu} \right]\left(\det\left[\Lambda(\tau) \right] \right)^{-1}\delta\left( S - k\cdot x_f\right).\nonumber
\end{eqnarray}
Here, we have used the invariance of the path-integral measure $\mathcal{D}p_{\Lambda}(\tau) = \mathcal{D}p(\tau)$ (see Appendix \ref{AppA}), since for
an element of $SO(1,d-2)$ we have $\det[\Lambda(\tau)]= 1$. We also made use of the global identity Eq.(\ref{eq_pint}). Thus, we can separate the remaining integrals in the form
\begin{eqnarray}
\Omega(S) &=& \frac{V_r}{C}  \int \mathcal{D}\Lambda \int d^{d}k\, \delta\left(m^2 + k^2\right) \delta\left( S - k\cdot x_f\right)\nonumber\\
&& \left[ \int \mathcal{D}p(\tau) \
\delta\left[p^{\mu}(\tau) - k^{\mu} \right] \right].
\label{VVO}
\end{eqnarray}
The path integral
in square brackets is evaluated by making use of the functional delta, to yield 
\begin{eqnarray}
\int \mathcal{D}p(\tau) \
\delta\left[p^{\mu}(\tau) - k^{\mu} \right] = 1.
\end{eqnarray}
Therefore, in Eq.(\ref{VVO})
we have, modulo an action independent normalization factor $\frac{V_r}{C}  \left[  \int_G \mathcal{D}\Lambda  \right]$ representing pure  redundancy,
that the desired phase-space volume of trayectories with equal action is given by
\bea
\Omega(S) = \int d^dk\ \delta\left(k^2 + m^2 \right) \delta\left(S -k_{\mu}x_f^{\mu}\right).
\label{OHJ}
\eea
Eq.\eqref{VVO} is a remarkable result: after factoring out the  redundancies associated to both reparametrizations and local velocity Lorentz transformations, the \textit{quantum} volume of paths of equal action that are physically inequivalent is equal to the \textit{classical} density of states $\Omega(S)$ of the Hamilton-Jacobi theory for the same action.

Now that $\Omega$ has been determined, it only remains to plug it back into \eqref{eq_2}, 
\bea
(x_f|0) &=& \mathcal{N}\int_{-\infty}^{\infty}dS\ e^{iS} \Omega(S)\\
&=& \mathcal{N}\int d^{d-1}k\int_{-\infty}^{\infty}dk_{0}\delta(-k_{0}^2 + \mathbf{k}^2 + m^2)\nonumber\\
&&\times\int_{-\infty}^{\infty}dS e^{i S}\delta(S - k\cdot x_f)\nonumber\\ \label{partiyeven}
&=& \mathcal{N}\int d^{d-1}k \left(\left.\frac{e^{ik\cdot x_f}}{2 \sqrt{\mathbf{k}^2 + m^2}} \right|_{k_{0} = \sqrt{\mathbf{k}^2+m^2}}\right.\\ \nonumber
&&\quad \quad+
\left.\left.\frac{e^{ik\cdot x_f}}{2 \sqrt{\mathbf{k}^2 + m^2}} \right|_{k_{0} = -\sqrt{\mathbf{k}^2+m^2}}\right)
.\nonumber
\eea
This is precisely the parity-even solution for the Klein-Gordon propagator $(x_f|0)= \Delta_1[x_f]$, 
as given for example in \cite{Itzykson:Book}.
It is exactly the parity {\it even} propagator since it was generated by paths which are all connected
by continuous transformations.
If one would like to obtain the parity-odd Klein-Gordon propagator, one would have to
modify the measure of the path integral, including virtual paths, 
that are connected to the two different sectors of the Lorentz group.
This technically complicated procedure can be circumvented
by simply changing one of the poles in the propagator (\ref{partiyeven}).

\section{Discussion}

The 
apparent incompatibility between the
relativistic propagator and unitarity. In non-relativistic quantum mechanics, the propagator fulfills
 the Chapman-Kolmogorov relation
 \be\label{Kol}
 (x_i|x_f)=\int d^dx_1(x_i|x_1)(x_1|x_f),
 \ee
where the integral is realized in spatial dimensions only. In contrast, 
the relativistic propagator  does not fulfill such a naive Chapman-Kolmogorov
relation, which is of course disturbing since it seems to indicate the collapse of
probability conservation. 
This inconsistency was noted by \cite{Prugo80,Fuku86,Kleinert:book}, who circumvented the problem
by turning to a phase space formulation or by introducing a spherical constraint.
It has also been argued that it is simply impossible to formulate
a probability conserving relativistic quantum mechanics and one has to go to quantum field theory right away.
Taking the problem more seriously it has also been argued that
the usual notion of probability has to be changed~\cite{Jizba:2008,Jizba:2010pi,Jizba:2011wg}.

However, those problems are solved when one realizes that most paths that appear in a naive
realization of the Chapman-Kolmogorov relation on the right hand side of (\ref{Kol}) are actually
equivalent through local velocity Lorentz transformations of the kind \eqref{dov}. 
They should not be integrated over and over again.
It is this type of overcounting which produces the seemingly non-conservation of 
probability in the path integral of the relativistic point particle.
Once, one takes into account this issue of equivalent intermediate steps,
the quantum propagation becomes unitary.
An explicit proof of this argument can 
be performed in a stepwise realization of (\ref{eq_2}), as shown in \cite{Koch:2017nha}.

\section{Conclusion}

This work is based on making notice that any term of the form $\dot{x}^2$ has a non-trivial symmetry of its own, as explained above.
Accounting for this symmetry allows one solve the practical problem of computing the path integral of the relativistic point particle (containing the square root) in a direct manner. 

We leave for a forthcoming paper the consequences that taking this larger symmetries into account might have in other systems such as Yang Mills, gravity, or string theory.

\section*{Acknowledgements}
We want to thank A. Faraggi and R. Abt, for helpful discussions. 
Thanks also to A. Canazas, A. Deriglazov, and A. Tolley for comments.
B.K. was supported by Fondecyt No 1161150,
E.M was supported by Fondecyt No 1141146,
I.R. was supported by CONICYT PCHA  2015149744.

\appendix

\section{Fadeev-Popov determinants}
\label{AppB}
Along the main body of the text, we have made use of a general functional identity
that is commonly used in the Fadeev-Popov \cite{Faddeev:1967fc} technique in Quantum Field Theory.
Let $\varphi(x)$ be a scalar field, and $\mathcal{F}[\varphi]$ a functional of these fields.
Then, the following integral identity holds \cite{Itzykson:Book2}
\bea
\int \mathcal{D}\varphi(x) \delta\left[\mathcal{F}[\varphi] \right] \det\left(\frac{\delta\mathcal{F}[\varphi(x)]}{\delta\varphi(x')}\right) = 1.
\label{eq_FadPop}
\eea

The determinant of the functional derivative that appears in the left-hand-side of Eq.(\ref{eq_FadPop}) is
commonly referred as the Fadeev-Popov determinant in the context of Abelian and Yang-Mills Field Theories  \cite{Itzykson:Book2,Faddeev:1967fc}.

\section{Change of integration variables}
\label{AppA}
In this appendix, we provide the details of the change of variables from
coordinates $x^{\mu}(\tau)$ to momenta $p^{\mu}(\tau)$.
In the continuum representation, we have defined the momenta
as derivatives with respect to the proper time,
\be
p^{\mu}(\tau) = m\frac{dx^{\mu}}{d\tau}.
\label{eq_A_p1}
\ee
In a  given discretization of the proper time, we have $d\tau \rightarrow \epsilon = (\tau_f - \tau_i)/M$, with
$M\rightarrow\infty$. Thus, each time step is defined as $\tau_k = \tau_i + k \epsilon$, with $0 \le k \le M$,
and the instantaneous coordinates become a set of discrete variables $x^{\mu}(\tau_k) \equiv x_{k}^{\mu}$. In the propagator,
the initial and final conditions are fixed as
\bea
x_{M}^{\mu} = x_f^{\mu}, \,\,\, x_{0}^{\mu} = 0,
\label{eq_const2}
\eea
and hence only $M-1$ coordinates $x_{k}^{\mu}$ are integrated along the trajectories.
The discrete,
finite differences version of Eq.(\ref{eq_A_p1}) is
\bea
p^{\mu}_{k} = m\frac{x^{\mu}_k - x^{\mu}_{k-1}}{\epsilon},\,\,\,1 \le k \le M
\label{eq_finite_diff}
\eea 
Despite Eq.(\ref{eq_finite_diff}) suggests that we have $M$ momenta, there exists a global constraint that reduces the total number of independent momenta
to $M-1$,
\bea
\sum_{k = 1}^{M} \epsilon\,p^{\mu}_{k} &=& m \sum_{k=1}^{M}\left(x_{k}^{\mu} - x_{k-1}^{\mu} \right)= m x_{f}^{\mu}.
\label{eq_const1}
\eea 
Here, in the second step we have applied the telescopic property of the sum.
In the continuum limit, Eq.(\ref{eq_const1}) becomes
\be
\int_{\tau_i}^{\tau_f} d\tau\, p^{\mu}(\tau) = m x_f^{\mu}.
\label{eq_const3}
\ee
The functional measure for the path-integral over space trajectories (notice that the
positions at $k=0$ and $k=M$ are fixed, by Eq.(\ref{eq_const2})) is defined as 
$\mathcal{D}x(\tau) = \prod_{k=1}^{M-1}\prod_{\mu=0}^{d-1}dx^{\mu}_{k}$. Therefore, the finite-differences Eq.(\ref{eq_finite_diff})
can be trivially inverted, to give a constant Jacobian:
\bea
\mathcal{D}x(\tau) &=& \prod_{k=1}^{M-1}\prod_{\mu=0}^{d-1}dx^{\mu}_{k} = \frac{\partial (x_1,\ldots,x_{M-1})}{\partial(p_1,\ldots,p_{M-1})} \prod_{k=1}^{M-1}\prod_{\mu=0}^{d-1}dp^{\mu}_{k}\nonumber\\
&=& \left(\frac{\epsilon}{m} \right)^{d(M-1)} \prod_{k=1}^{M-1}\prod_{\mu=0}^{d-1}dp^{\mu}_{k} \equiv \mathcal{D} p(\tau)
\label{eq_measure}
\eea
The momenta functional measure is invariant under local transformations of $SO(1,d-2)$, of the form $\Lambda_{\nu}^{\mu}(\tau)$ with $\det(\Lambda(\tau)) = 1$, such that
$p^{\mu}(\tau)\rightarrow \Lambda_{\nu}^{\mu}(\tau)p^{\nu}(\tau)\equiv p_{\Lambda}^{\mu}(\tau)$. Clearly, from the discrete definition of the measure Eq.(\ref{eq_measure}),
\bea
\mathcal{D}p_{\Lambda}(\tau) &=& \left(\frac{\epsilon}{m} \right)^{d(M-1)} \prod_{k=1}^{M-1}\prod_{\mu=0}^{d-1}d p_{\Lambda,k}^{\mu}\nonumber\\
&=& \left(\frac{\epsilon}{m} \right)^{d(M-1)} \prod_{k=1}^{M-1}\frac{\partial \left(p_{\Lambda,k}^{0},\ldots,p_{\Lambda,k}^{d-1}\right)}{\partial \left(p^{0}_{k},\ldots,p^{d-1}_k\right)}\prod_{\mu=0}^{d-1}dp^{\nu}_{k}\nonumber\\
&=& \left(\frac{\epsilon}{m} \right)^{d(M-1)} \prod_{k=1}^{M-1}\det[\Lambda_{k}]\prod_{\mu=0}^{d-1}dp^{\nu}_{k}\nonumber\\
&=& \mathcal{D}p(\tau)
\eea
where in the last two lines we have used the property that the Jacobian of the transformation $\frac{\partial \left(p_{\Lambda,k}^{0},\ldots,p_{\Lambda,k}^{d-1}\right)}{\partial \left(p^{0}_{k},\ldots,p^{d-1}_k\right)} = \det[\Lambda_{k}] = 1$.

\end{document}